\documentclass[conference]{IEEEtran}

\ifCLASSINFOpdf
\else
\fi
\usepackage{graphicx,graphics,epsfig,epstopdf,float}
\usepackage{amsmath}
\usepackage{amssymb}
\usepackage{mathrsfs}
\usepackage{enumerate}
\usepackage{subfigure}
\usepackage{colortbl}
\usepackage{color}
\usepackage{bm}
\usepackage{psfrag}
\usepackage{cite}
\usepackage{algorithm}
\usepackage{algorithmic}
\usepackage{mdwlist}
\usepackage{longtable}
\usepackage{array}
\usepackage{acronym}

\usepackage{multicol}
\usepackage{stfloats}
\usepackage{makecell}
\usepackage{epstopdf}

\usepackage{bm}

\DeclareMathAlphabet{\mathsfbr}{OT1}{cmss}{m}{n}
\SetMathAlphabet{\mathsfbr}{bold}{OT1}{cmss}{bx}{n}
\DeclareRobustCommand{\msf}[1]{%
  \ifcat\noexpand#1\relax\msfgreek{#1}\else\mathsfbr{#1}\fi
}

\makeatletter
\newcommand{\msfgreek}[1]{\csname s\expandafter\@gobble\string#1\endcsname}
\makeatother

\DeclareFontEncoding{LGR}{}{} 
\DeclareSymbolFont{sfgreek}{LGR}{cmss}{m}{n}
\SetSymbolFont{sfgreek}{bold}{LGR}{cmss}{bx}{n}
\DeclareMathSymbol{\salpha}{\mathord}{sfgreek}{`a}
\DeclareMathSymbol{\sbeta}{\mathord}{sfgreek}{`b}
\DeclareMathSymbol{\sgamma}{\mathord}{sfgreek}{`g}
\DeclareMathSymbol{\sdelta}{\mathord}{sfgreek}{`d}
\DeclareMathSymbol{\sepsilon}{\mathord}{sfgreek}{`e}
\DeclareMathSymbol{\szeta}{\mathord}{sfgreek}{`z}
\DeclareMathSymbol{\seta}{\mathord}{sfgreek}{`h}
\DeclareMathSymbol{\stheta}{\mathord}{sfgreek}{`j}
\DeclareMathSymbol{\siota}{\mathord}{sfgreek}{`i}
\DeclareMathSymbol{\skappa}{\mathord}{sfgreek}{`k}
\DeclareMathSymbol{\slambda}{\mathord}{sfgreek}{`l}
\DeclareMathSymbol{\smu}{\mathord}{sfgreek}{`m}
\DeclareMathSymbol{\snu}{\mathord}{sfgreek}{`n}
\DeclareMathSymbol{\sxi}{\mathord}{sfgreek}{`x}
\DeclareMathSymbol{\somicron}{\mathord}{sfgreek}{`o}
\DeclareMathSymbol{\spi}{\mathord}{sfgreek}{`p}
\DeclareMathSymbol{\srho}{\mathord}{sfgreek}{`r}
\DeclareMathSymbol{\ssigma}{\mathord}{sfgreek}{`s}
\DeclareMathSymbol{\stau}{\mathord}{sfgreek}{`t}
\DeclareMathSymbol{\supsilon}{\mathord}{sfgreek}{`u}
\DeclareMathSymbol{\sphi}{\mathord}{sfgreek}{`f}
\DeclareMathSymbol{\schi}{\mathord}{sfgreek}{`q}
\DeclareMathSymbol{\spsi}{\mathord}{sfgreek}{`y}
\DeclareMathSymbol{\somega}{\mathord}{sfgreek}{`w}

\DeclareMathSymbol{\svarsigma}{\mathord}{sfgreek}{`c}

\DeclareMathSymbol{\sGamma}{\mathalpha}{sfgreek}{`G}
\DeclareMathSymbol{\sDelta}{\mathalpha}{sfgreek}{`D}
\DeclareMathSymbol{\sTheta}{\mathalpha}{sfgreek}{`J}
\DeclareMathSymbol{\sLambda}{\mathalpha}{sfgreek}{`L}
\DeclareMathSymbol{\sXi}{\mathalpha}{sfgreek}{`X}
\DeclareMathSymbol{\sPi}{\mathalpha}{sfgreek}{`P}
\DeclareMathSymbol{\sSigma}{\mathalpha}{sfgreek}{`S}
\DeclareMathSymbol{\sUpsilon}{\mathalpha}{sfgreek}{`U}
\DeclareMathSymbol{\sPhi}{\mathalpha}{sfgreek}{`F}
\DeclareMathSymbol{\sPsi}{\mathalpha}{sfgreek}{`Y}
\DeclareMathSymbol{\sOmega}{\mathalpha}{sfgreek}{`W}

\DeclareRobustCommand{\mcal}[1]{%
  \ifcat\noexpand#1\relax\mathnormal{#1}\else\cal{#1}\fi
}
\DeclareRobustCommand{\BM}[1]{%
  \ifcat\noexpand#1\relax\bm{\boldUppercaseItalicGreek{#1}}\else\bm{#1}\fi
}
\makeatletter
\newcommand{\boldUppercaseItalicGreek}[1]{\csname var\expandafter\@gobble\string#1\endcsname}
\makeatother
\newcommand{\rv}[1]{\MakeLowercase{\msf{#1}}}

\newcommand{\V}[1]{\bm{#1}}

\DeclareMathAlphabet{\mathsfbr}{OT1}{cmss}{m}{n}
\SetMathAlphabet{\mathsfbr}{bold}{OT1}{cmss}{bx}{n}
\DeclareRobustCommand{\msf}[1]{%
  \ifcat\noexpand#1\relax\msfgreek{#1}\else\mathsfbr{#1}\fi
}

\DeclareFontEncoding{LGR}{}{} 
\DeclareSymbolFont{sfgreek}{LGR}{cmss}{m}{n}
\SetSymbolFont{sfgreek}{bold}{LGR}{cmss}{bx}{n}
\DeclareMathSymbol{\salpha}{\mathord}{sfgreek}{`a}
\DeclareMathSymbol{\sbeta}{\mathord}{sfgreek}{`b}
\DeclareMathSymbol{\sgamma}{\mathord}{sfgreek}{`g}
\DeclareMathSymbol{\sdelta}{\mathord}{sfgreek}{`d}
\DeclareMathSymbol{\sepsilon}{\mathord}{sfgreek}{`e}
\DeclareMathSymbol{\szeta}{\mathord}{sfgreek}{`z}
\DeclareMathSymbol{\seta}{\mathord}{sfgreek}{`h}
\DeclareMathSymbol{\stheta}{\mathord}{sfgreek}{`j}
\DeclareMathSymbol{\siota}{\mathord}{sfgreek}{`i}
\DeclareMathSymbol{\skappa}{\mathord}{sfgreek}{`k}
\DeclareMathSymbol{\slambda}{\mathord}{sfgreek}{`l}
\DeclareMathSymbol{\smu}{\mathord}{sfgreek}{`m}
\DeclareMathSymbol{\snu}{\mathord}{sfgreek}{`n}
\DeclareMathSymbol{\sxi}{\mathord}{sfgreek}{`x}
\DeclareMathSymbol{\somicron}{\mathord}{sfgreek}{`o}
\DeclareMathSymbol{\spi}{\mathord}{sfgreek}{`p}
\DeclareMathSymbol{\srho}{\mathord}{sfgreek}{`r}
\DeclareMathSymbol{\ssigma}{\mathord}{sfgreek}{`s}
\DeclareMathSymbol{\stau}{\mathord}{sfgreek}{`t}
\DeclareMathSymbol{\supsilon}{\mathord}{sfgreek}{`u}
\DeclareMathSymbol{\sphi}{\mathord}{sfgreek}{`f}
\DeclareMathSymbol{\schi}{\mathord}{sfgreek}{`q}
\DeclareMathSymbol{\spsi}{\mathord}{sfgreek}{`y}
\DeclareMathSymbol{\somega}{\mathord}{sfgreek}{`w}

\DeclareMathSymbol{\svarsigma}{\mathord}{sfgreek}{`c}

\DeclareMathSymbol{\sGamma}{\mathalpha}{sfgreek}{`G}
\DeclareMathSymbol{\sDelta}{\mathalpha}{sfgreek}{`D}
\DeclareMathSymbol{\sTheta}{\mathalpha}{sfgreek}{`J}
\DeclareMathSymbol{\sLambda}{\mathalpha}{sfgreek}{`L}
\DeclareMathSymbol{\sXi}{\mathalpha}{sfgreek}{`X}
\DeclareMathSymbol{\sPi}{\mathalpha}{sfgreek}{`P}
\DeclareMathSymbol{\sSigma}{\mathalpha}{sfgreek}{`S}
\DeclareMathSymbol{\sUpsilon}{\mathalpha}{sfgreek}{`U}
\DeclareMathSymbol{\sPhi}{\mathalpha}{sfgreek}{`F}
\DeclareMathSymbol{\sPsi}{\mathalpha}{sfgreek}{`Y}
\DeclareMathSymbol{\sOmega}{\mathalpha}{sfgreek}{`W}

\DeclareRobustCommand{\mcal}[1]{%
  \ifcat\noexpand#1\relax\mathnormal{#1}\else\cal{#1}\fi
}
\DeclareRobustCommand{\BM}[1]{%
  \ifcat\noexpand#1\relax\bm{\boldUppercaseItalicGreek{#1}}\else\bm{#1}\fi
}

\newtheorem{proposition}{Proposition}

\newtheorem{remark}{Remark}
\newtheorem{lemma}{Lemma}
\newcommand{\paperTitle}{Cooperative Vision-based Localization Networks with Communication Constraints}

\acrodef{twr}[TWR]{Two-Way ranging}%
\acrodef{tof}[ToF]{Time-of-Flight}%
\acrodef{sdstwr}[SDS-TWR]{symmetric double-sided TWR}%
\acrodef{dstwr}[DS-TWR]{double-sided TWR}%
\acrodef{alttwr}[AltDS-TWR]{alternative double-sided TWR}%
\acrodef{adstwr}[ADS-TWR]{asymmetric double-sided TWR}%
\acrodef{sstwr}[SS-TWR]{single-sided TWR}%

\acrodef{edtwr}[NB-DTWR]{Network Based Double TWR}%
\acrodef{dtwr}[D-TWR]{Double TWR}%
\acrodef{rtls}[RTLS]{real time location system}%
\acrodef{per}[PER]{Passive Extended Ranging}%
\acrodef{nbper}[NB-PR]{Network Based Passive Ranging}%
\acrodef{uwb}[UWB]{Ultra Wideband}%
\acrodef{ptd}[PTD]{Propagation Time Delay}%
\acrodef{iov}[IoV]{Internet of Vehicles}
\acrodef{gps}[GPS]{Global Positioning System}
\acrodef{fim}[FIM]{Fisher information matrix}
\acrodef{speb}[SPEB]{squared position error bound}
\acrodef{vgd}[V-GD]{variance-based gradient descent}
\acrodef{sa}[SA]{simulated annealing}
\acrodef{rmse}[RMSE]{root mean squared error}
\acrodef{toa}[TOA]{time of arrival}
\acrodef{nlos}[NLOS]{non-line-of-sight}
\acrodef{mec}[MEC]{multi-access edge computing}
\makeatletter
\renewcommand{\maketag@@@}[1]{\hbox{\m@th\normalsize\normalfont#1}}%
\makeatother

\begin{document}

\title{\paperTitle}

\author{
	\vspace{0.2cm}
	\IEEEauthorblockN{
		Fengzhuo~Zhang, Kai~Gu, and Yuan~Shen \\
	}
	\IEEEauthorblockA{
		Department of Electronic Engineering,
		Tsinghua University,
		Beijing 100084, China\\
		Beijing National Research Center for Information Science and Technology\\
	}
	Email: \{zfz15,
	guk16\}@mails.tsinghua.edu.cn,
	shenyuan\_ee@tsinghua.edu.cn
	
}

\maketitle

\begin{abstract}
Accurate location information is indispensable for the emerging applications of \ac{iov}, such as automatic driving and formation control. In the real scenario, vision-based localization has demonstrated superior performance to other localization methods for its stability and flexibility. In this paper, a scheme of cooperative vision-based localization with communication constraints is proposed. Vehicles collect images of the environment and distance measurements between each other. Then vehicles transmit the coordinates of feature points and distances with constrained bits to the edge to estimate their positions. The \ac{fim} for absolute localization is first obtained, based on which we derive the relative \ac{speb} through subspace projection. Furthermore, we formulate the corresponding bit allocation problem for relative localization. Finally, a \ac{vgd} algorithm is developed by considering the influence of photographing, distance measurements and quantization noises. Compared with conventional bit allocation methods, numerical results demonstrate the localization performance gain of our proposed algorithm with higher computational efficiency.
\end{abstract}

\acresetall

\section{Introduction}\label{sec:intro}
The past decade has witnessed the tremendous development of \ac{iov} in both fundamental theories and practical applications. As one of the most promising development directions, automation is the goal that \ac{iov} urgently pursues in the next generation. In the future, \ac{iov} will enable a number of applications for vehicle networks such as automatic driving, formation control and intelligent traffic management systems, where accurate and real-time position information is prerequisite for implementing high-level tasks \cite{8421288,8094945}. However, the \ac{gps}, which is usually used for localization in outdoor environment, tends to be incapable of providing reliable localization service for autonomous vehicles due to the high cost of deploying sufficient base stations. Moreover, localization signals from base stations are easily blocked by buildings around and interfered by other \ac{nlos} links, which further degrades the quality of localization service\cite{6427618}\cite{8422460}.

Visual localization is an emerging area of research that integrates 3D reconstruction techniques into network localization. To obtain a more accurate mapping of surroundings and locate sensors in harsh environment, intensive studies have been conducted to design robust visual localization algorithms to achieve the goals of environment reconstruction and localization simultaneously \cite{5979711,5940405}. These visual algorithms take the advantage that the visual observations received by cameras are not affected by multipath interference and can provide adequate position information to locate objects in the world coordinates. With the great advance in hardware and feature extraction algorithms, the computational efficiency of visual localization algorithms can be guaranteed to meet the increasing localization demand in GPS-denied environment \cite{5940405}. This paper investigates the point-based reconstruction algorithm for \ac{iov}, which extracts feature points from images and represents the scene with the point cloud \cite{5940405,Snavely:2006:PTE:1141911.1141964}. \par
\begin{figure}[t]
	\centering\includegraphics[width=0.9\columnwidth,draft=false]{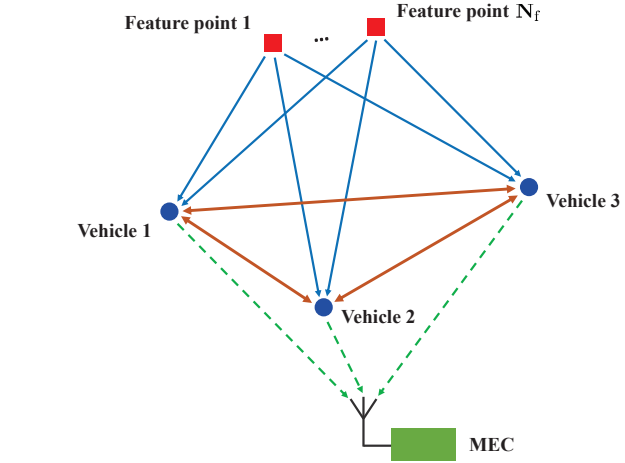}
	\caption{Cooperative reconstruction and localization in a mobile multi-vehicle network: each vehicle transmits the coordinates of feature points in the image and the distance measurements with limited bits to the edge.}
	\label{fig:scenario}
\end{figure}
To gain a better localization performance with limited observations, cooperation among sensors is profitable and deserves further investigating\cite{8421291}. In a real mobile scenario where the communication bandwidth between sensors is severely limited, it is impractical to transmit either all the feature points or the entire image to the multi-access edge computing (MEC) platform. Thus, an effective scheme of allocating bits among different vehicles and information sources is required for enhancing the visual localization performance under bandwidth constraints. However, little investigation has been carried out to introduce this kind of cooperative mechanism to visual localization.

In this paper, a bit allocation scheme is proposed for visual localization of vehicles. First, we introduce the system model of visual localization with communication constraints. Then the \ac{fim} of absolute positions of vehicles and feature points is derived, based on which the relative \ac{speb} is further given using the subspace projection method. On the basis of that, the bit allocation among different vehicles and measurements is formulated as an optimization problem by employing the metric of \ac{speb}. We propose a \ac{vgd} method to allocate bits with higher localization accuracy and computational efficiency compared to conventional algorithms.

\section{Problem Formulation}\label{sec:PF}
\subsection{Vision-based Localization Scheme}
Consider there are $N_{\rm v}$ vehicle, each equipped with a monocular camera (See Fig. \ref{fig:scenario}). To map the environment and derive the relative position of themselves, each vehicles uses the observed images to reconstruct the surroundings and measures the \ac{toa} to determine the distances between itself and others\cite{SheWin:J10a}. During an observation period, each camera extracts $N_{\rm f}$ most salient feature points in the current image, whose gradient is large enough so that feature points extracted by different vehicles can be shared among vehicles. Vehicles will transmit the coordinates of feature points and the measured distances between vehicles to the edge for estimating the relative positions of feature points and vehicles.\par
\subsection{System Model}
The absolute position of the \emph{i}th feature point is ${\mathbf{p}}_{i}=[x_{i}, y_{i}, z_{i}]^{\rm T}\in\mathbb{R}^{3}$ ($i=1,...,N_{\rm f}$). The augmented position vector of feature points is denoted by $\mathbf{p}=[\mathbf{p}_{1}^{\rm T},\mathbf{p}_{2}^{\rm T}, ... ,\mathbf{p}_{N_{\rm f}}^{\rm T}]^{\rm T}$. Similarly, the absolute position of the \emph{j}th vehicle is ${\mathbf{x}}_{j}=[x_{j}, y_{j}, z_{j}]^{\rm T}\in\mathbb{R}^{3}$ ($j=1,...,N_{\rm v}$) and the augmented position vector of vehicles is denoted by $\mathbf{x}=[\mathbf{x}_{1}^{\rm T}, \mathbf{x}_{2}^{\rm T}, ... ,\mathbf{x}_{N_{\rm v}}^{\rm T}]^{\rm T}$.
We assume that the camera on every vehicle has the same calibration matrix $\mathbf{K}$, whose elements represent the information about image resolution, coordinates of the principal point and the skew factor. Then the image of the feature point $\mathbf{p}_{i}$ at the \emph{j}th vehicle can be modeled as
\begin{align}
\left[\begin{array}{ccc} \mathbf{y}_{i} \\1\end{array}\right]=\frac{1}{\lambda_{ij}}\mathbf{K}\left [ \begin{array}{ccc}
{\mathbf{R}_{j}}^T, \ -{\mathbf{R}_{j}}^T\mathbf{x}_{j}\end{array}\right]\left[\begin{array}{ccc}\mathbf{p}_{i}\\ 1\end{array}\right]+
\left[\begin{array}{ccc}\rv{w}_{ij1}\\ \rv{w}_{ij2}\\0\end{array}\right]
\end{align}
where $\mathbf{y}_{i}$ is the coordinate vector of the feature point $\mathbf{p}_{i}$ in the image coordinate system. $\mathbf{R}_{j}$ indicates the Euler angles of the \emph{j}th vehicle, which can be obtained by the inertial measurement unit (IMU) on each vehicle. The noise terms of $\rv{w}_{ij1}$ and $\rv{w}_{ij2}$ represent the combined effect of photographing and quantization for X and Y coordinates, whose variances can be expressed as
\begin{align}
\sigma_{ijk}^2=\sigma_{ijk}^{\prime 2}+\sigma_{ijk}^{\prime\prime 2}, \quad k = 1,2\nonumber
\end{align}
where $\sigma_{ijk}^{\prime 2}$ denotes the variances of additive Gaussian noises produced by the photographing process of the \emph{i}th feature point at the \emph{j}th vehicle. Additionally, $\sigma_{ijk}^{\prime\prime 2}$ is the variances of quantization noises utilizing $b_{ijk}$ bits. Due to the independence between photographing noise and quantization noise\cite{1576972}, we can derive $\sigma_{ijk}^2$ by summing the above two parts. We further define the quantization bit allocation vector for X and Y coordinates as
\begin{align}
\mathbf{b}_{1}=[b_{111},b_{121},\ldots,b_{N_{\rm f}N_{\rm v}1},b_{112},\ldots,b_{N_{\rm f}N_{\rm v}2}]^{\rm T}.
\end{align}

Each pair of vehicles measure the distance between them based on \ac{toa}. The measurement $\rv{d}_{ij}$ between the \emph{i}th vehicle and the \emph{j}th vehicle can be modeled as
\begin{align}
\rv{d}_{ij}=\|\V{x}_{i}-\V{x}_{j}\|+\rv{W}_{ij3}
\end{align}
where $\rv{W}_{ij3}$ denotes the combination of measurement and quantization noises. The corresponding noise variance can be written as
\begin{align}
\sigma_{ij3}^2=\sigma_{ij3}^{\prime 2}+\sigma_{ij3}^{\prime\prime 2}\nonumber
\end{align}
among which $\sigma_{ij3}^{\prime 2}$ is the measurement noise variance while $\sigma_{ij3}^{\prime\prime 2}$ accounts for the quantization effect. We define the bit allocation vector for measured distances as
\begin{align}
\mathbf{b}_{2}=[b_{123},b_{133},\cdots,b_{(N_{\rm v}-1)N_{\rm v}3}]^{\rm T}.
\end{align}
 We adopt a probabilistic quantization method to quantize the coordinates of feature points and the distance between vehicles\cite{4156398}. We suppose that the observed signal is bounded to $[0,2W]$, i.e. $x=\theta+n\in[0,2W]$. $W$ is decided by the physical constraints of parameter $\theta$.  We first divide $[0,2W]$ into $2^{b}-1$ equilong intervals with the spacing $\Delta=\frac{2W}{(2^{b}-1)}$. If $n\Delta \leq x \leq (n+1)\Delta$ for $0\leq n\leq 2^{b}-2$, then $x$ is quantized to $\hat{x}(b)$ with $b$ bits as
 \begin{align}
 &P(\hat{x}(b)=n\Delta) = 1-\frac{x-n\Delta}{\Delta} \notag \\
 &P(\hat{x}(b)=(n+1)\Delta) = \frac{x-n\Delta}{\Delta}.\nonumber
 \end{align}
 It can be proved that $\hat{x}(b)$ is an unbiased estimation of $\theta$ and the variance of noise satisfies\cite{4156398}
\begin{align}
\mathbb{E}\{|\hat{x}(b)-\theta|^2\}\leq \sigma^2+\frac{W^2}{(2^b-1)^2},\quad \forall b\in  \mathbb{N}_{+}\nonumber
\end{align}
where $\sigma^2$ is the variance of observation noise and $\frac{W^2}{(2^b-1)^2}$ is introduced as quantization noise.\par
To guarantee the localization performance with communication constraints, we assume $\sigma_{ijk}^{\prime\prime 2}=\frac{W_{k}^2}{(2^{b_{ijk}}-1)^2}$ for $ k=1,2,3$ where $W_{1}\times W_{2}$ is the resolution of images and $W_{3}$ is the size of the scene which we set in advance.\par
To facilitate the following performance analysis in terms of \ac{fim}, we further assume that the quantization noise follows a  Gaussian distribution as $\mathcal{N}(0,\sigma_{ijk}^{\prime\prime 2})$\cite{1413469}.
\subsection{Performance Metric}
In order to measure the performance of vision-based localization, we first need to determine a tractable metric for 3D reconstruction. However, a great number of works have explored this topic but fail to reach a consensus. Among them, point-based algorithms are frequently utilized as practical solutions to this issue \cite{5979711}, \cite{5940405}, \cite{Snavely:2006:PTE:1141911.1141964}. To evaluate the accuracy of 3D reconstruction, we employ the point-to-point distance as the reconstruction performance metric, which is widely used in image registration \cite{924423}, \cite{121791}, given by
\begin{align}
\epsilon=\sum_{i}\|\mathbf{p}_{i}-\mathbf{\hat{p}}_{i}\|^2
\end{align}
where $\mathbf{p}_{i}$ is the actual position vector of the $i$th feature point and $\mathbf{\hat{p}}_{i}$ is the reconstructed position vector. On that basis, we derive the overall performance metric for joint 3D reconstruction and vehicle localization with communication constraints in the following sections.
\subsection{Relative Localization}
We denote the position vector of feature points and vehicles as $\widetilde{\mathbf{p}}=[\mathbf{p}^{\rm T}, \mathbf{x}^{\rm T}]^{\rm T}\in\mathbb{R}^{3(N_{\rm f}+N_{v})}$. Then the transformation of the estimated position vector $\hat{\widetilde{\mathbf{p}}}$ can be defined as
\begin{align}
T_{\bm{\alpha}}(\hat{\widetilde{\mathbf{p}}})&=\hat{\widetilde{\mathbf{p}}}+x\mathbf{v}_{x}+y\mathbf{v}_{y}+z\mathbf{v}_{z} \notag \\
\mathbf{v}_{x}&=[1,0,0,1,0,0,...,1,0,0]^{\rm T}\in\mathbb{R}^{3(N_{\rm f}+N_{v})} \notag \\
\mathbf{v}_{y}&=[0,1,0,0,1,0,...,0,1,0]^{\rm T}\in\mathbb{R}^{3(N_{\rm f}+N_{v})} \notag \\
\mathbf{v}_{z}&=[0,0,1,0,0,1,...,0,0,1]^{\rm T}\in\mathbb{R}^{3(N_{\rm f}+N_{v})}
\end{align}
where $\bm{\alpha}=[x,y,z]^{\rm T}$ is the transformation parameter. The optimal transformation parameter can be defined as
\begin{align}
\bm{\alpha}_{0}=\mathop{\arg\min}_{\bm{\alpha}} \ \ \| \widetilde{\mathbf{p}}-T_{\bm{\alpha}}(\hat{\widetilde{\mathbf{p}}}) \|.
\end{align}
The optimal solution can be derived as
\begin{align}
\bm{\alpha}_{0}=\frac{(\widetilde{\mathbf{p}}-\hat{\widetilde{\mathbf{p}}})^{\rm T}[\mathbf{v}_{x}, \mathbf{v}_{y}, \mathbf{v}_{z}]}{\mathbf{v}_{x}^{\rm T}\mathbf{v}_{x}}.
\end{align}
The estimated position transformed by $\bm{\alpha}_{0}$ can be written as $\hat{\widetilde{\mathbf{p}}}_{0}=T_{\bm{\alpha}_{0}}(\hat{\widetilde{\mathbf{p}}})$. Then the  total error $\epsilon$ can be expressed as the sum of two parts, i.e.,
\begin{eqnarray}
\epsilon = \epsilon_{t}+\epsilon_{r}
\end{eqnarray}
where $\epsilon_{t} = \|[\mathbf{v}_{x}, \mathbf{v}_{y}, \mathbf{v}_{z}]\bm{\alpha}_{0}\|^{2}$ and $\epsilon_{r} = \|\widetilde{\mathbf{p}}-\hat{\widetilde{\mathbf{p}}}_{0}\|^{2}$ denote the transformation error and the relative error, respectively. Since no position information is acquired from anchors, we focus on the derivation and analysis of the relative error in this paper.
\subsection{Performance Bound}
In addition to approximating the distribution of the noise term $\rv{w}_{ijk}$ by $N(0,\sigma_{ijk}^2)$, we further define the augmented bit allocation vector as ${\bf b}=[{\bf b}_1^{\rm T}, {\bf b}_2^{\rm T}]^{\rm T}$. Then we derive the \ac{fim} of parameters $\widetilde{\mathbf{p}}$ as follow \cite{SheWymWin:J10}
\begin{align}
\mathbf{J}(\widetilde{\mathbf{p}},{\bf b})=\mathbf{J}_{1}(\widetilde{\mathbf{p}},{\bf b}_1)+\mathbf{J}_{2}(\widetilde{\mathbf{p}},{\bf b}_2).
\end{align}

The first term of $\mathbf{J}_{1}(\widetilde{\mathbf{p}},{\bf b}_1)$ is generated by the observation of feature points, which can be written as
\begin{align}
\mathbf{J}_{1}(\widetilde{\mathbf{p}},{\bf b}_1) =\left[ \begin{array}{ccc}\mathbf{A} \qquad& \mathbf{B}\\\mathbf{B}^{\rm T}\qquad&\mathbf{C}\end{array} \right].
\end{align}
The corresponding submatrices have the form
\begin{align}
\mathbf{A}&={\rm diag} \bigg\{ \sum_{j=1}^{N_{\rm v}}\mathbf{G}_{1j},\sum_{j=1}^{N_{\rm v}}\mathbf{G}_{2j},\cdots,\sum_{j=1}^{N_{\rm v}}\mathbf{G}_{N_{\rm f}j} \bigg\} \\
\mathbf{B}&=
\left[ \begin{array}{cccc}
-\mathbf{G}_{11}\ &-\mathbf{G}_{12}\ &\cdots & -\mathbf{G}_{1N_{\rm v}}\\
-\mathbf{G}_{21}\ &-\mathbf{G}_{22}\ &\cdots & -\mathbf{G}_{2N_{\rm v}}\\
\vdots\ &\vdots\ & \ddots & \vdots\\
-\mathbf{G}_{N_{\rm f}1}\ &-\mathbf{G}_{N_{\rm f}2}\ &\cdots &-\mathbf{G}_{N_{\rm f}N_{\rm v}}
\end{array} \right]\\
\mathbf{C}&= {\rm diag} \bigg\{ \sum_{i=1}^{N_{\rm f}}\mathbf{G}_{i1},\sum_{i=1}^{N_{\rm f}}\mathbf{G}_{i2},\cdots,\sum_{i=1}^{N_{\rm f}}\mathbf{G}_{iN_{\rm v}} \bigg\}
\end{align}
where
\begin{align}
\mathbf{G}_{ij}
&=\sum_{k=1}^2 \frac{(f_{ij3}\mathbf{v}_{kj}-f_{ijk}\mathbf{v}_{3j})(f_{ij3}\mathbf{v}_{kj}-f_{ijk}\mathbf{v}_{3j})^{\rm T}}{\sigma_{ijk}^{2}} \\
f_{ijk}&=\frac{\mathbf{v}_{kj}^{\rm T}(\mathbf{p}_{i}-\mathbf{x}_{j})}{[\mathbf{v}_{3j}^{\rm T}(\mathbf{p}_{i}-\mathbf{x}_{j})]^{2}}, \quad k=1,2,3
\end{align}
with $\mathbf{v}_{1j}, \mathbf{v}_{2j}$ and $\mathbf{v}_{3j}$ as the row vectors of $\mathbf{K}{\mathbf{R}_{j}^{\rm T}}$, i.e.,
\begin{align}
\mathbf{K}{\mathbf{R}_{j}^{\rm T}} = [ \mathbf{v}_{1j}, \mathbf{v}_{2j}, \mathbf{v}_{3j} ]^{\rm T}.
\end{align}

The second term of $\mathbf{J}_{2}(\widetilde{\mathbf{p}},{\bf b}_2)$ represents the information from distance measurements between vehicles, given by
\begin{align}
\mathbf{J}_{2}(\widetilde{\mathbf{p}},{\bf b}_2) & =\left [ \begin{array}{ccc} \mathbf{0}\qquad&\mathbf{0} \\
\mathbf{0}^{\rm T}\qquad&\mathbf{D}\end{array}\right]
\end{align}
where
\begin{align}
\mathbf{D} & =\left [ \begin{array}{cccc}
\sum_{j}\mathbf{S}_{1j}\ &-\mathbf{S}_{12}\quad &\cdots &-\mathbf{S}_{1N_{\rm v}}\\
-\mathbf{S}_{21}\ &\sum_{j}\mathbf{S}_{2j}\quad &\cdots &-\mathbf{S}_{2N_{\rm v}}\\
\vdots\ &\vdots & \ddots & \vdots\\
-\mathbf{S}_{N_{\rm v}1}\ &-\mathbf{S}_{N_{\rm v}2} & \cdots &\sum_{j}\mathbf{S}_{N_{\rm v}j}\\
\end{array}\right]
\end{align}
and
\begin{align}
\mathbf{S}_{ij}&=\frac{1}{\sigma_{ij3}^2}\mathbf{w}_{ij}\mathbf{w}_{ij}^{\rm T}\\
\mathbf{w}_{ij}&=\frac{\mathbf{x}_{i}-\mathbf{x}_{j}}{\|\mathbf{x}_{i}-\mathbf{x}_{j}\|}.
\end{align}
As we only concern about the relative positions of vehicles and feature points, we can simply determine the "shape" of the position of vehicles and feature points. It is proved that $\mathbf{J}(\widetilde{\mathbf{p}},{\bf b})$ is rank-deficient and can be decomposed as \cite{4542553}
\begin{align}
\mathbf{J}(\widetilde{\mathbf{p}},{\bf b})=[\mathbf{U}\ \mathbf{\widetilde{U}}] \left[ \begin{array}{cc}\mathbf{\Lambda} & \mathbf{0}\\ \mathbf{0}^{\rm T} & \mathbf{0}\end{array} \right][\mathbf{U}\ \mathbf{\widetilde{U}}]^{\rm T}
\end{align}
where $\mathbf{\Lambda}$ is the diagonal matrix whose diagonal elements are the nonzero eigenvalues of $\mathbf{J}(\widetilde{\mathbf{p}},{\bf b})$. $\mathbf{U}$ and $\mathbf{\widetilde{U}}$ are comprised of eigenvectors corresponding to nonzero and zero eigenvalues, respectively. Note that $\mathbf{U}$ captures all the relative position information while $\widetilde{\mathbf{U}}$ provides no extra information for the determination of relative errors. The rank of $\widetilde{\mathbf{U}}$ is 3 in most circumstances\footnote{A network consisting of two vehicles and one feature point will lead to a special case of $\mathbf{\widetilde{U}}$ whose rank is 4.}, given by
\begin{align}
\mathbf{\widetilde{U}} = \left[ \, \frac{\mathbf{v}_{x}}{\|\mathbf{v}_{x}\|},\ \frac{\mathbf{v}_{y}}{\|\mathbf{v}_{y}\|},\ \frac{\mathbf{v}_{z}}{\|\mathbf{v}_{z}\|} \, \right].
\end{align}
Then we can derive the relative \ac{speb} as
\begin{align}
\rm{P}_{r}(\widetilde{\mathbf{p}},{\bf b})&=\rm{trace}\{(\mathbf{U}(\mathbf{U}^{\rm T}\mathbf{J}(\widetilde{\mathbf{p}},{\bf b})\mathbf{U})^{-1}\mathbf{U}^{\rm T})\}
\end{align}
and we will employ it as the performance metric for optimization in the following section.
\section{Bit Allocation Algorithms}
\subsection{Optimization Problem Formulation}
In this subsection, we formulate the bit allocation problems for visual localization. The goal of bit allocation is to achieve the minimum relative \ac{speb} given a limited total bit number $B$ for vehicles. The problem is given as
\begin{align}
\mathscr{P}:\quad \min\limits_{ \{b_{ijk}\} } \quad &\rm{P}_{r}(\widetilde{\mathbf{p}},{\bf b}) \notag \\
{\rm s.t.} \quad &\sum_{i,j,k} b_{ijk}\leq B \notag, \quad b_{ijk} \ge 0.
\end{align}
\begin{lemma}
	$f(x)=\dfrac{1}{1+\frac{a^{2}}{(2^{x}-1)^{2}}}$ is concave when $x\geq \log_{2}a$ for $a\gg 1$.\par
\end{lemma}
\begin{IEEEproof}
We first calculate the second derivative of $f(x)$ with respect to $x$ as
\begin{align}
f^{\prime\prime}(x)&=2a^2\ln 2\bigg\{\frac{[2^{x}(2^{x}-1)\ln 2+2^{2x}\ln 2]}{[a^2+(2^{x}-1)^{2}]^{2}} \notag \\
&\quad -\frac{2^{2x+2}(2^{x}-1)^{2}\ln 2}{[a^2+(2^{x}-1)^{2}]^{3}} \bigg\}.
\end{align}
Let $y=2^{x}$, then
\begin{align}
f^{\prime\prime}(x)=0\Leftrightarrow 2y^{3}-3y^{2}-2a^{2}y+(a^{2}+1)=0.
\end{align}
According to Cardano formula, the equation has three real roots and the first root can be written as
\begin{align}
y_{1}&=\sqrt[3]{-\frac{q}{2}+\sqrt{\left(\frac{q}{2}\right)^2+\left(\frac{p}{3}\right)^3}}+\sqrt[3]{-\frac{q}{2}-\sqrt{\left(\frac{q}{2}\right)^2+\left(\frac{p}{3}\right)^3}}
\end{align}
where
$p=-a^{2}-3/4 \quad q=1/4$. For $a\gg 1$, we have
\begin{align}
y_{1}\approx 2\cdot\frac{\sqrt{3}}{2}\sqrt[6]{\left(\frac{q}{2}\right)^{2}-\left(\frac{q}{2}\right)^{2}-\left(\frac{p}{3}\right)^{3}}
\approx \sqrt{3}\frac{a}{\sqrt{3}}=a.
\end{align}
Similarly, we have $y_{2}\approx -a$ and $y_{3}\approx 0$. Since $\lim\limits_{x\to\infty}f^{\prime\prime}(x)\le 0$, then
$f^{\prime\prime}(x)\le 0$ when $x \ge \log_{2}{a}$, i.e., $y\ge a$.
So $f(x)$ is concave when $x\geq \log_{2}a.$
\end{IEEEproof}
\begin{remark}
In our problem, we choose the value of $a$ as ${W_{k}}/{\sigma_{ijk}^{\prime }}$. For $k=1,2$, $W_{k}$ tends to be 1024 or 768 and $\sigma_{ijk}^{\prime}$ is 40. For $k=3$, $W_{k}$ tends to be 300 while $\sigma_{ijk}^{\prime}$ is about 4. ${W_{k}}/{\sigma_{ijk}^{\prime}} \gg 1$ holds in our setting.
\end{remark}
\begin{proposition}
When $b_{ijk}\geq \log_{2} ({W_{k}}/{\sigma_{ijk}^\prime})$, the relative \ac{speb} $\rm{P}_{r}(\widetilde{\mathbf{p}},{\bf b})$ is convex with respect to $b_{ijk}$.
\end{proposition}
\begin{IEEEproof}
We define $g: \mathbb{N}^{2N_{\rm v}N_{\rm f}+\frac{N_{\rm v}(N_{\rm v}-1)}{2}} \!\rightarrow\! \mathbb{S}^{3N_{\rm v}+3N_{\rm f}}$ as
\begin{align}
g(\mathbf{b})=\mathbf{J}(\widetilde{\mathbf{p}},{\bf b}).
\end{align}
From Lemma 1, $g(\mathbf{b})$ is K-concave when $b_{ijk}\geq \log_{2} ({W_{k}}/{\sigma_{ijk}^{\prime}})$. For $\mathbf{J}(\widetilde{\mathbf{p}},{\bf b}) \succeq 0$, $[\mathbf{J}^{-1}(\widetilde{\mathbf{p}},{\bf b})]_{m,m}$ is a convex and non-increasing function with respect to ${\bf b}$ \cite{4012317}. Thus, $\rm{P}_{r}(\widetilde{\mathbf{p}},{\bf b})$ is convex with respect to $b_{ijk}$ when $b_{ijk}\geq \log_{2} ({W_{k}}/{\sigma_{ijk}^{\prime}})$.
\end{IEEEproof}

\subsection{Variance-based Gradient Descent Algorithm}
The objective function of bit allocation is non-convex due to the nonlinearity of $f(x)$. The complexity of the brute force algorithm is too high to implement. The method of \ac{sa} is an alternative sub-optimal algorithm, which needs to verify hundreds of trial solutions. In this subsection, we present a \ac{vgd} method, which requires much less computation time than the \ac{sa} approach but can achieve better performance  when the bit number is larger than the number of measurements.

To make full use of the acquired feature point and position information, we allocate more bits to those vehicles with more accurate observations. The accuracy of the measurement depends on two factors: the range $W_{k}$ and the variance of the observation noise $\sigma_{ijk}^{\prime2}$. The measurement which is bounded in a shorter interval will provide more information than the measurement bounded in a longer interval when allocated with the same number of bits. The measurement whose variance of observation noise is smaller will contain more information. However, allocating all bits to the vehicle with the smallest observation noise and range could not  guarantee that the \ac{fim} is positive definite and the total variance of this measurement decreases little when allocated with more bits. For this reason, we start with allocating bits among all the measurements and then adopt an iterative algorithm to minimize $\rm{P}_{r}$.

To set the initial solution, we need to allocate bits among the coordinates of feature points and the distances between vehicles. We first determine the ratio $m$ between the number of bits allocated to the feature points and the distances among vehicles by grid search among $[0,1]$. With a fixed ratio $m$, we then derive the initial solution by distributing bits among different points or distances proportionally to $1/(\sigma_{ijk}^{\prime}\log_{2}{W_{k}})$. We take the logarithm of $W_{k}$ because it is the numerator of $\frac{W_{k}^2}{(2^{b_{ijk}}-1)^2}$. \par
Then we adopt the gradient descent (GD) algorithm to find the optimal solution in an iterative manner. When $b_{ijk}$ exceeds a threshold, we use the steepest descent (SD) algorithm to accelerate the search process instead. The last step is to discretize the allocation bit vector since the GD and SD algorithms will generate the non-integer bit allocation solution. As it is inefficient to search all the possible integer  solutions, we randomly allocate the sum of the fractional parts for a number of times, followed by taking the one which has the minimum value of $\rm{P}_{r}$ as the final allocation strategy.\par
\begin{algorithm}[t]
	\caption{Variance-based Gradient Descent Algorithm} \label{dla2}
	\textbf{Input:} $\sigma_{ijk}^{\prime}$, $\delta$, $B$\\
	\textbf{Output:} $b_{ijk}$ \\
	\textbf{Procedure:}
	\begin{algorithmic}[1]
		\FOR{$m =$ 0 to 1}
		\STATE $m=m+\delta$
		\vspace{0.1 cm}
		\STATE $b_{ijk} = \frac{ mB/(\sigma_{ijk}^{\prime}\log_2W_k) }
		{\sum\limits_{i,j} \big( 1/(\sigma_{ij1}^{\prime}\log_2W_1) + 1/(\sigma_{ij2}^{\prime}\log_2W_2) \big)}, k=1,2$
		\vspace{0.2 cm}
		\STATE $b_{ij3} = \frac{ (1-m)\textbf{}B/(\sigma_{ij3}^{\prime}\log_2W_3) }{\sum\limits_{i,j} \big( 1/(\sigma_{ij3}^{\prime}\log_2W_3) \big)}$
		\vspace{0.2 cm}
		\REPEAT
		\STATE $\rm{P}_{\rm r}^*=\rm{P}_{\rm r}$
		\IF{$b_{ijk}\ge \log_2({{W_k}/{\sigma_{ijk}^{\prime}}})$}
		\STATE $k=\mathop{\arg\min}_{k} \rm{P}_{r}$
		\ELSE
		\STATE Generate $k$ from $\mathcal{U}(0,1)$
		\ENDIF
		\STATE $\mathbf{b}=\mathbf{b}+k\nabla_{\mathbf{b}}{\rm{P}_{r}}$
		\STATE Update $\rm{P}_{r}$ with ${\bf b}$
		\vspace{0.1 cm}
		\UNTIL {$\Big|\frac{\rm{P}_{r}-\rm{P}_{\rm r}^*}{\rm{P}_{\rm r}^*}\Big|\leq 10^{-5}$}
		\vspace{0.1 cm}
		\ENDFOR
  		\STATE Select ${\bf b}^*$ from  $\lfloor{1}/{\delta}\rfloor$ alternative solutions
  		\STATE Discretize ${\bf b}^*$ to derive the optimal allocation vector
	\end{algorithmic}
\end{algorithm}
\par
\subsection{Decoupling Optimization Algorithm}
Optimizing the number of bits allocated to all the measurements simultaneously demands too much time as the problem is not convex in the entire feasible domain. A natural alternative is to first optimize the bit allocation among the measurements of the same feature point, then optimize the allocation among the measurements from the same vehicle. The similar process can be implemented to optimize the bit allocation among distance measurements. Since the number of bits allocated to feature points or to distances is invariable during the optimization, we adopt the grid search algorithm to find the suboptimal ratio between these two parts.\par
\subsection{Simulated Annealing Algorithm}
The \ac{sa} algorithm is a general algorithm to find the optimal solution of a non-convex problem. It typically includes initialization, generation of a new solution and Metropolis algorithm. We set the initial solution by allocating bits randomly. The process of generating a new solution is conducted by adding one bit to a measurement and subtracting one bit from another measurement randomly.
\begin{algorithm}[t]
	\caption{Decoupling Optimization Algorithm} \label{decoup}
	\textbf{Input:} $\sigma_{ijk}^2$, $\delta$, $B$\\
	\textbf{Output:} $b_{ijk}$ \\
	\textbf{Procedure:}
	\begin{algorithmic}[1]
		\FOR{$m =$ 0 to 1}
		\STATE $m=m+\delta$
		\vspace{0.1 cm}
		\STATE $b_{ijk} = \frac{mB}
		{2N_{\rm v}N_{\rm f}},\ k=1,2$
		\vspace{0.2 cm}
		\STATE $b_{ij3} = \frac{2(1-m)B}{N_{\rm v}(N_{\rm v}-1)}$
		\vspace{0.2 cm}
		\STATE Transform $\mathbf{b}$ into two matrices $\mathbf{A}_{N_{\rm f}\times 2N_{\rm v}}$, $\mathbf{B}_{N_{\rm v}\times N_{\rm v}}$
		\STATE $N=0$
		\REPEAT
		\STATE Optimize $\mathbf{A}$ row by row and column by column
		\STATE Optimize $\mathbf{B}$ row by row and column by column
		\STATE $N=N+1$
		\UNTIL {$N\ge N_{0}$}
		\STATE Transform $\mathbf{A}$, $\mathbf{B}$ into $\mathbf{b}$
		\ENDFOR
		\STATE Select ${\bf b}^*$ from $\lfloor{1}/{\delta}\rfloor$ alternative solutions
		\STATE Discretize ${\bf b}^*$ to derive the optimal allocation vector
	\end{algorithmic}
\end{algorithm}
\begin{remark}
The above algorithms use precise position knowledge of the vehicle network to derive the relative \ac{speb}. In our future work, the uncertainty of position parameters will be taken into account for implementing robust optimization.
\end{remark}
\section{Numerical Results}\label{sec:NR}
In this section, we present numerical results for the proposed bit allocation method. The simulation scenario is a square region [-25m, 25m] $\times$ [-25m, 25m]. Five vehicles are uniformly placed at a circle whose radius is 5m. We set $W_{3}$ as 250m. Seventy feature points are randomly placed in the 5m $\times$ 5m $\times$ 2m cuboid whose center locates at origin. The resolution of images we use in the simulation is $3264\times 2488$. The scale of the sensor in cameras is 36mm $\times$ 23.9mm. The focal length of cameras is $600$mm.\par
We compare our \ac{vgd} algorithm with the uniform allocation scheme that assigns the total bits equally over all the measurements as well as the \ac{sa} algorithm.\par
\begin{figure}[t]
	\centering\includegraphics[width=0.9\columnwidth,draft=false]{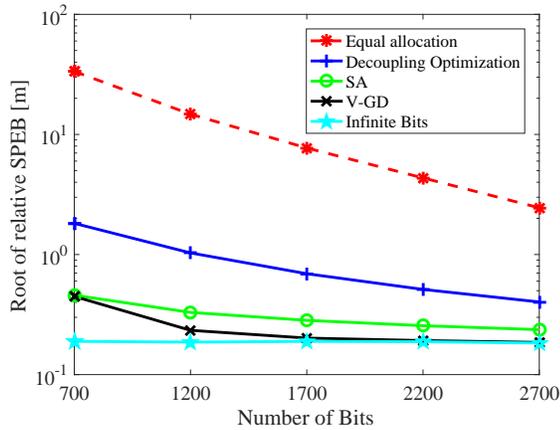}
	\caption{The root of relative \acp{speb} with respect to the number of bits.}
	\label{fig:simu1}
\end{figure}

Fig. \ref{fig:simu1} shows the \acp{speb} as a function of $B$. It can be observed that the performance of all the algorithms reaches the relative \ac{speb} with infinite bits. The equal allocation strategy has the poorest performance because it treats all the observations with different noise variances fair. The \ac{speb} of decoupling optimization algorithm decreases slowly as it does not take the coupling relation of different images and distance measurements into account. In our simulation, $\mathbf{b}$ is a 710 dimensional vector, which is too high for the \ac{sa} to find the optimal solution given the non-convex optimization problem. The design of the initial point of the \ac{vgd} algorithm makes it feasible to converge to the optimal solution, which is indicated by the lower relative \ac{speb} than that of the \ac{sa}. It can be seen that the  relative \ac{speb} can achieve the ideal performance bound with infinite bits when $B$ is larger than 1700.\par
\begin{figure}[t]
	\centering\includegraphics[width=0.9\columnwidth,draft=false]{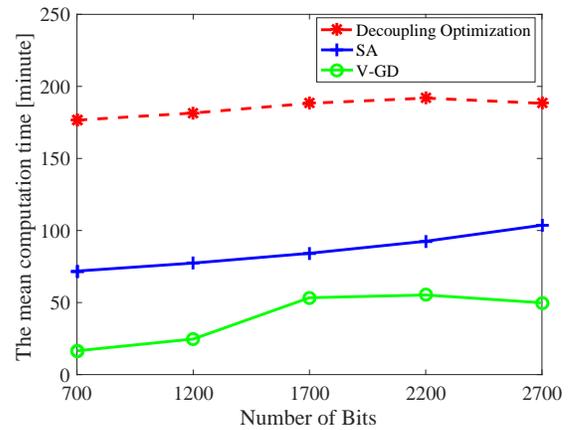}
	\caption{The mean computation time with respect to the number of bits.}
	\label{fig:simu2}
\end{figure}

In Fig. \ref{fig:simu2}, the mean computation time of two suboptimal bit allocation algorithms are compared with the \ac{vgd}. In every iteration of the decoupling optimization algorithm, we optimize over a $N_{\rm f}$ or $N_{\rm v}$ dimensional vector, so its mean time is higher than that of the \ac{vgd} algorithm. The computation time of \ac{sa} depends on the product of the initial temperature and the number of iterations at each temperature. To avoid the convergence to some local optimal solutions, the initial temperature of the \ac{sa} should be set high enough and a sufficient number of iterations should be carried out at each temperature. Conversely, the \ac{vgd} algorithm only needs to implement the GD until it converges owing to its proper initialization. It can be seen that the \ac{vgd} algorithm reduces the mean time by around $50\%$ compared with the \ac{sa}. The mean computation time of \ac{vgd} decreases a little on the range of $1700$ to $2700$ for the reason that the optimization problem is more likely to be convex as the number of bits increases. Since the number of feature points is larger than that in the realistic situation, \ac{vgd} only takes several seconds in the setting of realistic situation.
\section{Conclusion}\label{sec:Con}
In this paper, we developed a bit allocation scheme for vision-based localization in vehicles networks. The absolute \ac{speb} for feature points and vehicles was first derived. We then formulated the optimization problem for bit allocation in terms of relative \ac{speb}. The local convexity of the objective function was proved. Based on that, a \ac{vgd} algorithm was proposed. Numerical results show that the \ac{vgd} algorithm outperforms the \ac{sa} and the decoupling algorithms. Meanwhile, the \ac{vgd} algorithm reduces the computation time by half compared with the \ac{sa} algorithm. Our work demonstrates the potential of cooperative vehicle networks and provides a solution to bit allocation for high-accuracy vision-based relative localization. In the future, we will investigate the influence of  the mobility of vehicles and the bit allocation strategy in a harsh communication environment.
\section*{Acknowledgment}
This research was supported by the National Natural Science Foundation of China under Grant 61871256 and 61811530329.

%



%



\acresetall		




%

\bibliographystyle{IEEEtran}

\bibliography{IEEEabrv,StringDefinitions,SGroupDefinition,SGroup}

\ifCLASSOPTIONcaptionsoff
  \newpage
\fi
 \newpage

\end{document}